\documentclass[runningheads]{llncs}
\usepackage[T1]{fontenc}
\usepackage{indentfirst}
\usepackage{microtype}
\usepackage{cite}
\usepackage{amsmath,amssymb}
\usepackage{graphicx}
\usepackage{booktabs}
\usepackage{array}
\usepackage{tabularx}
\usepackage{url}
\usepackage{xcolor}
\usepackage{tcolorbox}
\usepackage{tikz}
\usetikzlibrary{arrows.meta}
\usepackage{pgfplots}
\usepackage{pgfplotstable}
\pgfplotsset{compat=1.18}
\usepackage[hidelinks]{hyperref}
\newcommand{\PaperTitle}{Large Language Models for Quantum Code Translation: An Exploratory Study}

\hypersetup{pdftitle={\PaperTitle},pdfauthor={Xiyuan Li, Yuechen Li, Jianjun Zhao}}
\newcommand{\pass}[1]{\mbox{Pass@#1}}
\newif\ifshowrevisions
\showrevisionsfalse
\colorlet{RevisionColor}{black}

\DeclareRobustCommand{\revised}[1]{%
  \ifshowrevisions\textcolor{RevisionColor}{#1}\else#1\fi}
\DeclareRobustCommand{\unmarked}[1]{{\normalcolor#1}}

\newif\ifshowreviewmarkup
\showreviewmarkupfalse
\DeclareRobustCommand{\ReviewChange}[1]{%
  \ifshowreviewmarkup\textcolor{red}{#1}\else#1\fi}

\newcommand{\StudyValue}[1]{%
  \ifcsname study@#1\endcsname
    \csname study@#1\endcsname
  \else
    \PackageError{study-values}{Unknown study value: #1}%
      {Run Python on scripts/update-study.py after updating data/.}%
  \fi}
\expandafter\def\csname study@assertions.categories.count\endcsname{5}
\expandafter\def\csname study@assertions.category-list\endcsname{format/dimensions, output constraints, probability/distribution, numerical accuracy, and operator equivalence}
\expandafter\def\csname study@assertions.events.all\endcsname{298}
\expandafter\def\csname study@assertions.events.baseline\endcsname{199}
\expandafter\def\csname study@assertions.items\endcsname{77}
\expandafter\def\csname study@assertions.pl.events\endcsname{105}
\expandafter\def\csname study@assertions.pl.numerical_accuracy.share\endcsname{19.0}
\expandafter\def\csname study@assertions.pl.operator_equivalence.share\endcsname{20.0}
\expandafter\def\csname study@assertions.pl.output_format.share\endcsname{0.0}
\expandafter\def\csname study@assertions.pl.output_validity.share\endcsname{30.5}
\expandafter\def\csname study@assertions.pl.probability_distribution.share\endcsname{30.5}
\expandafter\def\csname study@assertions.qs.events\endcsname{94}
\expandafter\def\csname study@assertions.qs.numerical_accuracy.share\endcsname{10.6}
\expandafter\def\csname study@assertions.qs.operator_equivalence.share\endcsname{17.0}
\expandafter\def\csname study@assertions.qs.output_format.share\endcsname{0.0}
\expandafter\def\csname study@assertions.qs.output_validity.share\endcsname{23.4}
\expandafter\def\csname study@assertions.qs.probability_distribution.share\endcsname{48.9}
\expandafter\def\csname study@assertions.samples.all\endcsname{241}
\expandafter\def\csname study@augmented.slots\endcsname{1,720}
\expandafter\def\csname study@baseline.failed\endcsname{1,787}
\expandafter\def\csname study@baseline.passed\endcsname{1,653}
\expandafter\def\csname study@baseline.slots\endcsname{3,440}
\expandafter\def\csname study@conditions.augmented\endcsname{2}
\expandafter\def\csname study@conditions.count\endcsname{6}
\expandafter\def\csname study@delta.glm.pl.1\endcsname{18.37}
\expandafter\def\csname study@delta.glm.pl.3\endcsname{16.55}
\expandafter\def\csname study@delta.glm.qs.1\endcsname{10.93}
\expandafter\def\csname study@delta.glm.qs.3\endcsname{9.67}
\expandafter\def\csname study@delta.max\endcsname{18.37}
\expandafter\def\csname study@delta.min\endcsname{7.67}
\expandafter\def\csname study@delta.pro.pl.1\endcsname{7.67}
\expandafter\def\csname study@delta.pro.pl.3\endcsname{6.88}
\expandafter\def\csname study@delta.pro.qs.1\endcsname{8.14}
\expandafter\def\csname study@delta.pro.qs.3\endcsname{7.64}
\expandafter\def\csname study@efficiency.flash.output-tokens\endcsname{8,541}
\expandafter\def\csname study@efficiency.flash.runtime\endcsname{92.00}
\expandafter\def\csname study@efficiency.glm.output-tokens\endcsname{9,527}
\expandafter\def\csname study@efficiency.glm.runtime\endcsname{148.30}
\expandafter\def\csname study@efficiency.gpt.output-tokens\endcsname{2,758}
\expandafter\def\csname study@efficiency.gpt.runtime\endcsname{149.84}
\expandafter\def\csname study@efficiency.pro.output-tokens\endcsname{7,426}
\expandafter\def\csname study@efficiency.pro.runtime\endcsname{108.17}
\expandafter\def\csname study@evaluation.seed\endcsname{42}
\expandafter\def\csname study@evaluation.timeout_seconds\endcsname{180}
\expandafter\def\csname study@example.family\endcsname{less\_than}
\expandafter\def\csname study@failures.assertion_failed\endcsname{160}
\expandafter\def\csname study@failures.assertion_failed.share\endcsname{9.0}
\expandafter\def\csname study@failures.candidate_execution_error\endcsname{143}
\expandafter\def\csname study@failures.candidate_execution_error.share\endcsname{8.0}
\expandafter\def\csname study@failures.delivery-compilation.share\endcsname{74.1}
\expandafter\def\csname study@failures.missing_generation_artifact\endcsname{700}
\expandafter\def\csname study@failures.missing_generation_artifact.share\endcsname{39.2}
\expandafter\def\csname study@failures.sdk_compatibility_error\endcsname{160}
\expandafter\def\csname study@failures.sdk_compatibility_error.share\endcsname{9.0}
\expandafter\def\csname study@failures.sdk_compilation_error\endcsname{624}
\expandafter\def\csname study@failures.sdk_compilation_error.share\endcsname{34.9}
\expandafter\def\csname study@failures.with_program_prompt.assertion_failed\endcsname{81}
\expandafter\def\csname study@failures.with_program_prompt.assertion_failed.share\endcsname{8.9}
\expandafter\def\csname study@failures.with_program_prompt.candidate_execution_error\endcsname{98}
\expandafter\def\csname study@failures.with_program_prompt.candidate_execution_error.share\endcsname{10.8}
\expandafter\def\csname study@failures.with_program_prompt.missing_generation_artifact\endcsname{333}
\expandafter\def\csname study@failures.with_program_prompt.missing_generation_artifact.share\endcsname{36.7}
\expandafter\def\csname study@failures.with_program_prompt.sdk_compatibility_error\endcsname{46}
\expandafter\def\csname study@failures.with_program_prompt.sdk_compatibility_error.share\endcsname{5.1}
\expandafter\def\csname study@failures.with_program_prompt.sdk_compilation_error\endcsname{349}
\expandafter\def\csname study@failures.with_program_prompt.sdk_compilation_error.share\endcsname{38.5}
\expandafter\def\csname study@failures.with_program_prompt.total\endcsname{907}
\expandafter\def\csname study@failures.without_program_prompt.assertion_failed\endcsname{53}
\expandafter\def\csname study@failures.without_program_prompt.assertion_failed.share\endcsname{4.8}
\expandafter\def\csname study@failures.without_program_prompt.candidate_execution_error\endcsname{79}
\expandafter\def\csname study@failures.without_program_prompt.candidate_execution_error.share\endcsname{7.2}
\expandafter\def\csname study@failures.without_program_prompt.missing_generation_artifact\endcsname{492}
\expandafter\def\csname study@failures.without_program_prompt.missing_generation_artifact.share\endcsname{44.7}
\expandafter\def\csname study@failures.without_program_prompt.sdk_compatibility_error\endcsname{70}
\expandafter\def\csname study@failures.without_program_prompt.sdk_compatibility_error.share\endcsname{6.4}
\expandafter\def\csname study@failures.without_program_prompt.sdk_compilation_error\endcsname{407}
\expandafter\def\csname study@failures.without_program_prompt.sdk_compilation_error.share\endcsname{37.0}
\expandafter\def\csname study@failures.without_program_prompt.total\endcsname{1,101}
\expandafter\def\csname study@figure.prompt\endcsname{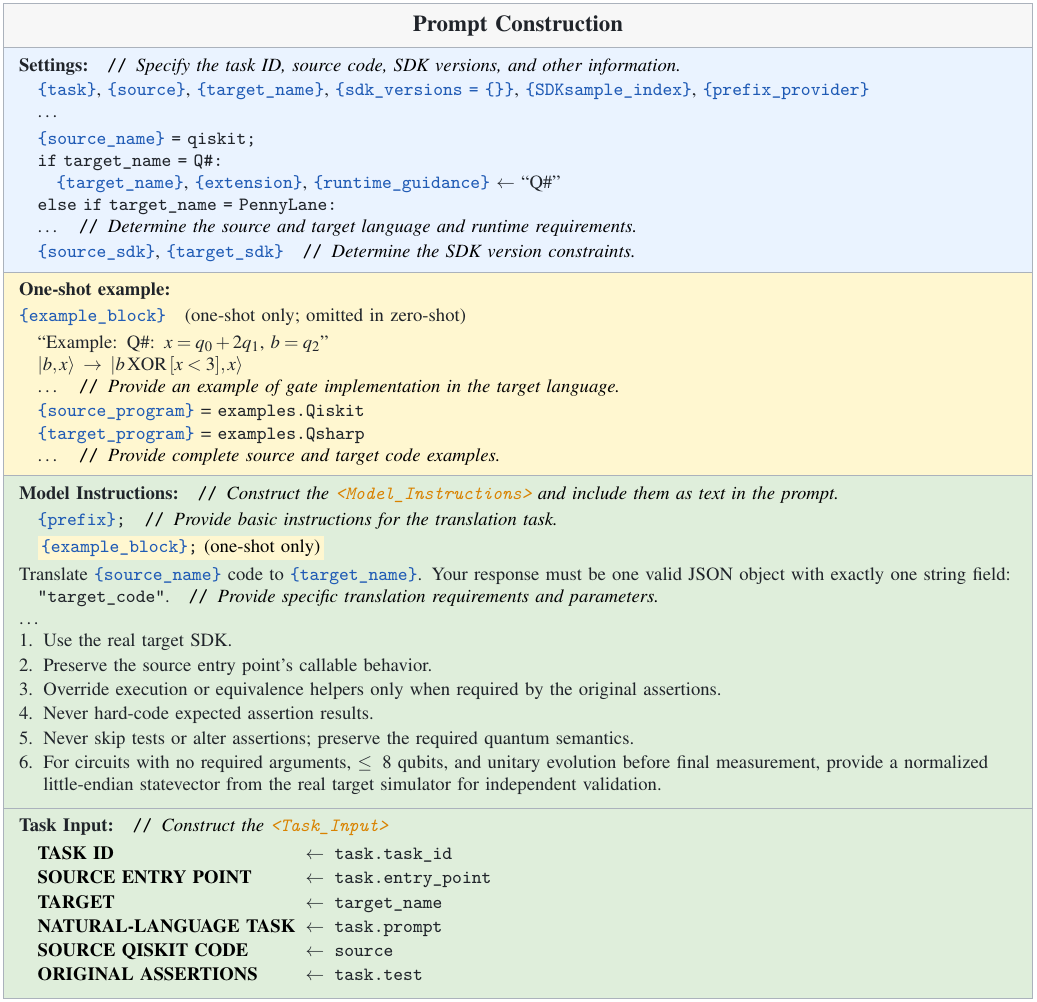}
\expandafter\def\csname study@figure.prompt.width\endcsname{0.88}
\expandafter\def\csname study@figure.workflow\endcsname{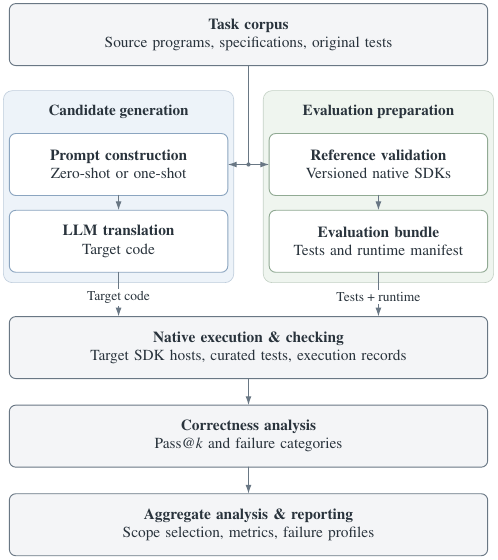}
\expandafter\def\csname study@figure.workflow.width\endcsname{0.72}
\expandafter\def\csname study@generation.timeout\endcsname{180}
\expandafter\def\csname study@generation.tokens\endcsname{12,000}
\expandafter\def\csname study@metric.flash.pl.1\endcsname{72.33}
\expandafter\def\csname study@metric.flash.pl.3\endcsname{86.53}
\expandafter\def\csname study@metric.flash.qs.1\endcsname{44.88}
\expandafter\def\csname study@metric.flash.qs.3\endcsname{65.19}
\expandafter\def\csname study@metric.glm.pl.1\endcsname{49.30}
\expandafter\def\csname study@metric.glm.pl.3\endcsname{65.43}
\expandafter\def\csname study@metric.glm.qs.1\endcsname{22.56}
\expandafter\def\csname study@metric.glm.qs.3\endcsname{48.97}
\expandafter\def\csname study@metric.gpt.pl.1\endcsname{51.63}
\expandafter\def\csname study@metric.gpt.pl.3\endcsname{61.36}
\expandafter\def\csname study@metric.gpt.qs.1\endcsname{71.63}
\expandafter\def\csname study@metric.gpt.qs.3\endcsname{82.11}
\expandafter\def\csname study@metric.ks\endcsname{1,3}
\expandafter\def\csname study@metric.pro.pl.1\endcsname{55.81}
\expandafter\def\csname study@metric.pro.pl.3\endcsname{73.26}
\expandafter\def\csname study@metric.pro.qs.1\endcsname{16.28}
\expandafter\def\csname study@metric.pro.qs.3\endcsname{37.71}
\expandafter\def\csname study@model.flash.gateway\endcsname{openrouter}
\expandafter\def\csname study@model.flash.model\endcsname{deepseek/deepseek-v4.1-flash}
\expandafter\def\csname study@model.flash.name\endcsname{DeepSeek-V4.1-Flash}
\expandafter\def\csname study@model.flash.reasoning_plan_effort\endcsname{high}
\expandafter\def\csname study@model.flash.samples\endcsname{10}
\expandafter\def\csname study@model.flash.seed.first\endcsname{42}
\expandafter\def\csname study@model.flash.seed.last\endcsname{51}
\expandafter\def\csname study@model.flash.temperature\endcsname{0.8}
\expandafter\def\csname study@model.flash.timeout\endcsname{180}
\expandafter\def\csname study@model.flash.tokens.final\endcsname{3,000}
\expandafter\def\csname study@model.flash.tokens.plan\endcsname{9,000}
\expandafter\def\csname study@model.flash.tokens.total\endcsname{12,000}
\expandafter\def\csname study@model.glm.gateway\endcsname{openrouter}
\expandafter\def\csname study@model.glm.model\endcsname{z-ai/glm-5.2}
\expandafter\def\csname study@model.glm.name\endcsname{GLM-5.2}
\expandafter\def\csname study@model.glm.reasoning_effort\endcsname{high}
\expandafter\def\csname study@model.glm.samples\endcsname{10}
\expandafter\def\csname study@model.glm.seed.first\endcsname{42}
\expandafter\def\csname study@model.glm.seed.last\endcsname{51}
\expandafter\def\csname study@model.glm.temperature\endcsname{0.8}
\expandafter\def\csname study@model.glm.timeout\endcsname{180}
\expandafter\def\csname study@model.glm.tokens.total\endcsname{12,000}
\expandafter\def\csname study@model.gpt.gateway\endcsname{openai}
\expandafter\def\csname study@model.gpt.model\endcsname{gpt-5.6-sol}
\expandafter\def\csname study@model.gpt.name\endcsname{GPT-5.6-sol}
\expandafter\def\csname study@model.gpt.reasoning_effort\endcsname{medium}
\expandafter\def\csname study@model.gpt.samples\endcsname{10}
\expandafter\def\csname study@model.gpt.timeout\endcsname{180}
\expandafter\def\csname study@model.gpt.tokens.total\endcsname{12,000}
\expandafter\def\csname study@model.pro.gateway\endcsname{openrouter}
\expandafter\def\csname study@model.pro.model\endcsname{deepseek/deepseek-v4-pro}
\expandafter\def\csname study@model.pro.name\endcsname{DeepSeek-V4-Pro}
\expandafter\def\csname study@model.pro.reasoning_plan_effort\endcsname{high}
\expandafter\def\csname study@model.pro.samples\endcsname{10}
\expandafter\def\csname study@model.pro.seed.first\endcsname{42}
\expandafter\def\csname study@model.pro.seed.last\endcsname{51}
\expandafter\def\csname study@model.pro.temperature\endcsname{0.8}
\expandafter\def\csname study@model.pro.timeout\endcsname{180}
\expandafter\def\csname study@model.pro.tokens.final\endcsname{6,000}
\expandafter\def\csname study@model.pro.tokens.plan\endcsname{6,000}
\expandafter\def\csname study@model.pro.tokens.total\endcsname{12,000}
\expandafter\def\csname study@models.closed-source\endcsname{GPT-5.6-sol}
\expandafter\def\csname study@models.count\endcsname{4}
\expandafter\def\csname study@models.open-weight\endcsname{GLM-5.2, DeepSeek-V4-Pro, and DeepSeek-V4.1-Flash}
\expandafter\def\csname study@rq.1.question\endcsname{How does zero-shot quantum code translation performance vary across model configurations and target SDKs?}
\expandafter\def\csname study@rq.1.title\endcsname{Zero-shot Functional Correctness across Model Configurations and Target SDKs}
\expandafter\def\csname study@rq.2.question\endcsname{How do correctness and failure profiles differ between zero-shot and one-shot prompts?}
\expandafter\def\csname study@rq.2.title\endcsname{Translation Correctness under Zero-shot and One-shot Prompting}
\expandafter\def\csname study@rq.3.question\endcsname{What are the main failure categories in LLM-based quantum code translation?}
\expandafter\def\csname study@rq.3.title\endcsname{Failure Categories in Quantum Code Translation}
\expandafter\def\csname study@samples.per-task\endcsname{10}
\expandafter\def\csname study@samples.total\endcsname{5,160}
\expandafter\def\csname study@sdk.pennylane\endcsname{0.45.1}
\expandafter\def\csname study@sdk.qdk\endcsname{1.32.0}
\expandafter\def\csname study@sdk.qiskit\endcsname{1.4.6}
\expandafter\def\csname study@sdk.qiskit-aer\endcsname{0.17.2}
\expandafter\def\csname study@targets.count\endcsname{2}
\expandafter\def\csname study@tasks\endcsname{43}
\expandafter\def\csname study@tasks.excluded\endcsname{38}
\expandafter\def\csname study@tasks.original\endcsname{44}

\newcommand{\RQQuestion}[1]{\StudyValue{rq.#1.question}}

\newcommand{\PromptField}[1]{\texttt{\{\{}\nolinkurl{#1}\texttt{\}\}}}
\newcommand{\StudyIdentifier}[1]{%
  \begingroup
  \edef\StudyIdentifierText{\StudyValue{#1}}%
  \expandafter\nolinkurl\expandafter{\StudyIdentifierText}%
  \endgroup}

\usepackage{ifthen}

\newboolean{commentsenabled}
\setboolean{commentsenabled}{true}

\ifthenelse{\boolean{commentsenabled}}{
	\newcommand{\nbc}[3]{%
		{\colorbox{#3}{\bfseries\sffamily\scriptsize\textcolor{white}{#1}}}%
		{\textcolor{#3}{\sf\small$\langle$\textit{#2}$\rangle$}}%
	}
	
}{
	\newcommand{\nbc}[3]{}

}

\begin{document}
\title{\PaperTitle}
\titlerunning{LLMs for Quantum Code Translation}
\author{Xiyuan Li\inst{1} \and Yuechen Li\inst{2} \and
Jianjun Zhao\inst{1}\thanks{Corresponding author: Jianjun Zhao.}}
\authorrunning{X. Li et al.}
\institute{Kyushu University\\
\email{li.xiyuan.868@s.kyushu-u.ac.jp, zhao@ait.kyushu-u.ac.jp}
\and
Beihang University\\
\email{liyuechen@buaa.edu.cn}}
\maketitle
\begin{abstract}
Quantum software development is challenged by quantum-mechanical semantics and fragmented software development kits (SDKs), which complicate cross-SDK code translation and execution. In this paper, we propose an LLM-based approach for translating executable programs between high-level quantum SDKs.
We develop a quantum-aware translation workflow whose prompt design and validation reflect quantum semantics and accommodate the version evolution of quantum ecosystems.
We conduct a preliminary evaluation of \StudyValue{models.count} LLM configurations on \StudyValue{tasks} tasks, translating Qiskit programs to PennyLane and Q\#.
LLMs achieve a zero-shot \pass{1} of up to \StudyValue{metric.flash.pl.1}\%, while one-shot prompting further improves the translation effectiveness. These results suggest potential for quantum code translation by LLMs.
\keywords{Quantum software engineering \and Code translation \and Large language models}
\end{abstract}

\section{Introduction}


Quantum computing has attracted growing interest, leading to a diverse ecosystem of software development kits (SDKs) and programming languages, including Qiskit~\cite{qiskit-docs}, PennyLane~\cite{pennylane-docs}, and Q\#~\cite{qsharp-docs}.
However, quantum software development remains challenging because quantum-mechanical abstractions, such as superposition, entanglement, and probabilistic measurement, are unfamiliar to many software engineers. Meanwhile, heterogeneous SDK interfaces and backend-specific execution models require developers to reconcile different application programming interfaces (APIs) and runtime constraints, complicating quantum software development and maintenance.

Code translation~\cite{chen2025systematic,jiang2026multilingual} is a natural software-engineering response to this interoperability problem.
For example, a controlled-$X$ gate can be expressed in Qiskit as
\texttt{qc.cx(a, b)} and in Q\# as
\texttt{Controlled X([q[a]], q[b]);}, where \texttt{a} and \texttt{b}
index the control and target qubits, respectively.
A practical translator should perform this API mapping while preserving the source program's intended behavior or code structure. 
Large language models (LLMs) are promising candidates for this task because they can interpret code and natural-language specifications, generate code in different programming languages, and have supported activities in quantum software engineering, such as code generation~\cite{quanbench}, comment generation~\cite{yu2025quainth}, and static analysis~\cite{shin2025quantum}.
Such intuition and observations motivate us to investigate whether LLMs can translate quantum programs across heterogeneous SDKs while preserving their required behavior and structure.


To the best of our knowledge, this paper presents the first proposal of LLM-based quantum code translation between high-level SDKs or programming languages to deal with developer-facing quantum programming.
We propose a quantum-aware LLM translation workflow from input tasks to translated programs along with the aggregate analysis. 
In particular, we contribute a prompt-construction method tailored to quantum code translation: it captures quantum-mechanical semantics, such as qubit ordering and the roles of control and target qubits, together with SDK-specific syntax, version constraints, runtime requirements, and task specifications, thereby accommodating the rapid evolution of quantum programming frameworks.
We preliminarily evaluate four LLM configurations on 43 benchmark tasks, translating Qiskit programs to PennyLane and Q\# and comparing zero-shot with one-shot prompting for two models. The best zero-shot \pass{1} reaches 72.33\% on PennyLane and 71.63\% on Q\#, while one-shot prompting can substantially improve \pass{1}. 
Furthermore, we find that missing source artifacts and SDK compilation errors dominate zero-shot failures, accounting for 39.2\% and 34.9\%, respectively. These findings suggest the potential of LLM-based quantum code translation while identifying key barriers for future research.

\section{Background and Related Work}


\textbf{LLM-based translation of classical code.} LLMs have been applied to translate classical code across programming languages~\cite{lost-translation}. UniTrans uses execution feedback~\cite{unitrans}, InterTrans explores intermediate translation paths~\cite{intertrans}, and AlphaTrans decomposes repositories and combines translation with multiple validation levels~\cite{alphatrans}.
These advances motivate applying LLMs to quantum code translation, where preserving behavior also requires accounting for relative phase, bit ordering, control and target roles, and probabilistic measurement outcomes.

\textbf{Quantum code migration across versions.}
Prior work investigates how quantum code can be migrated across SDK versions~\cite{suarez2025automatic}, which involves refactoring the original code to accommodate changes in APIs. By contrast, quantum code translation transfers programs across heterogeneous SDKs or programming languages, which requires reconciling different programming abstractions and execution conventions, rather than handling incremental updates within the same SDK.

\textbf{Quantum code transpilation.}
Transpilation in quantum computing refers to the process of transforming a quantum circuit into an equivalent one that is compatible with the target quantum hardware or simulator. Several works also perform cross-SDK transpilation, such as Siavash et al. suggesting the use of LLMs~\cite{siavash2025llm} and LinguaQuanta leveraging the gate decomposition~\cite{linguaquanta}. However, quantum code transpilation is rooted in the hardware-oriented conversion in terms of compilers and backends, while our quantum code translation targets the high-level SDKs that could be used by developers to design quantum code. 


\textbf{Quantum code generation.} 
Quantum code generation studies use model-driven engineering or LLMs, including (1) model-based generation, like QuanUML~\cite{guo2025quanuml} and M2QCode~\cite{guo2025m2qcode}; (2) LLM benchmarking studies like Qiskit Code Assistant~\cite{qiskit-assistant}, PennyCoder~\cite{pennycoder}, QuanBench~\cite{quanbench}, and QuanBench+~\cite{quanbench-plus}; and (3) LLM agents such as QPipe~\cite{tao2026leveraging} and QAgent~\cite{fu2025qagent}. Compared to code generation, including multi-platform generation for one task~\cite{guo2025m2qcode}, our translation task converts a given source program into a target program in another SDK or language while preserving functionality and general structure.


\section{Translation Workflow}
\label{sec:workflow}

\subsection{Workflow Overview}

\begin{figure}[htbp]
\centering
\includegraphics[width=0.72\linewidth]{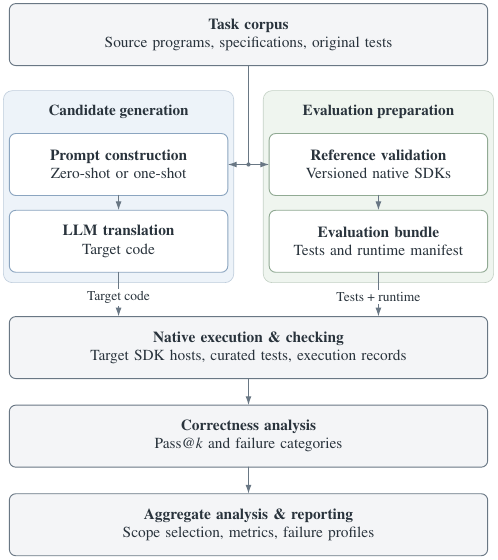}
\caption{Workflow for LLM-based quantum code translation.}
\label{fig:workflow}
\end{figure}

Figure~\ref{fig:workflow} summarizes the translation workflow. \textbf{Candidate generation} uses the source program and task specification to produce a target program. In parallel, \textbf{evaluation preparation} comprises two blocks. Reference validation checks reference implementations against the tests in versioned native SDK environments. \ReviewChange{The Evaluation bundle block} packages the validated tests and a runtime manifest for subsequent evaluation. The generated program is then \textbf{executed and checked} against this evaluation bundle.
Execution results feed into \textbf{correctness analysis}, which distinguishes failures to deliver or execute a candidate from assertion rejections and relates the latter to the properties tested. \textbf{Aggregate analysis and reporting} summarizes these outcomes across tasks, models, and target SDKs (although Q\# is a programming language, we use ``SDKs'' as shorthand for PennyLane and the Q\# development environment to improve readability in both the paper and the experimental code).

Within the workflow, prompt construction is central to quantum code translation because it defines the task information and translation constraints provided to the model. The next subsection describes how these prompts are built. 

\subsection{Prompt Construction}
\label{sec:workflow-generation}
\begin{figure}[!htb]
\centering
\makebox[\linewidth][c]{\includegraphics[width=0.88\linewidth]{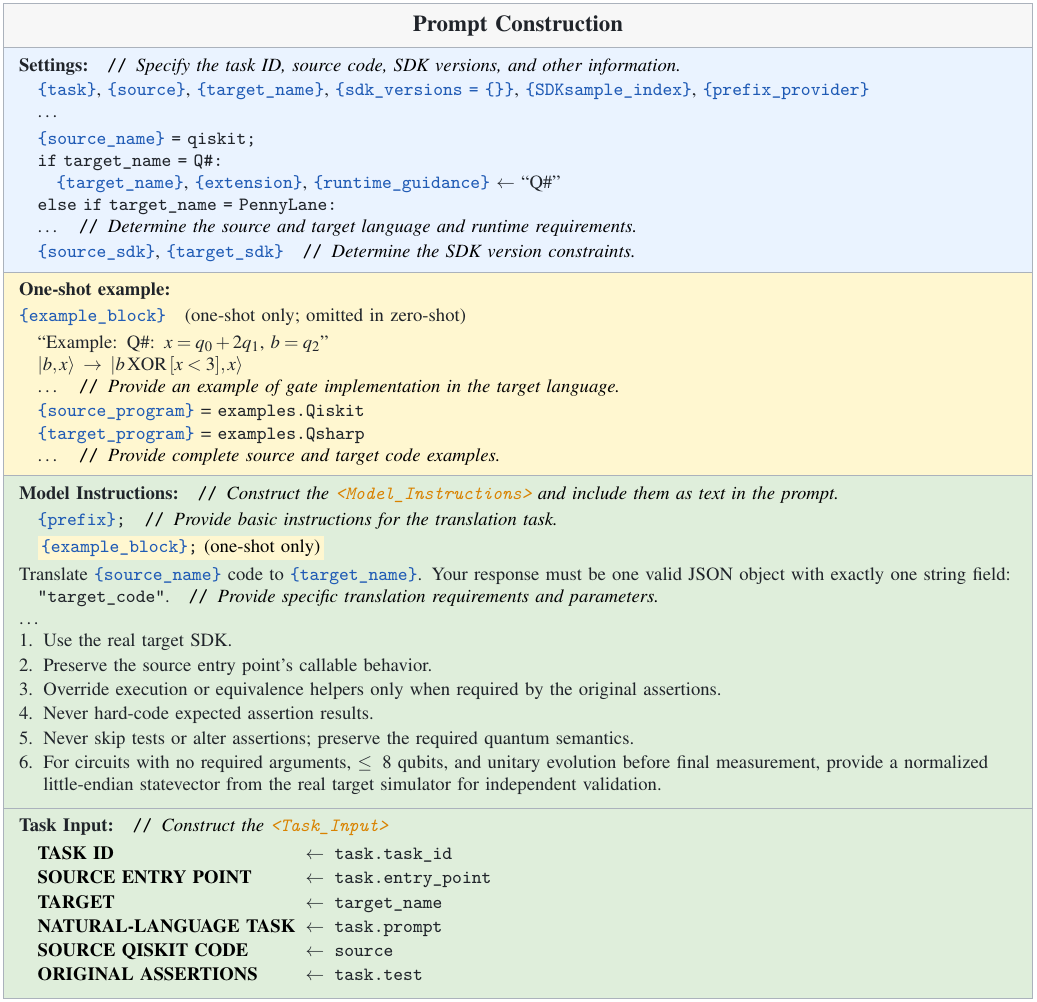}}
\caption{Prompt construction process and content used in the translation experiments.}
\label{fig:one-shot-program-prompt}
\end{figure}

Figure~\ref{fig:one-shot-program-prompt} illustrates the prompt construction process and prompt content used in our translation experiments. \textbf{Black text} denotes literal prompt fields; \textbf{blue text} denotes general placeholders; \textit{italic text} provides brief comments for the prompt; and \textbf{orange text} highlights two core prompt components to specify the translation task.
Specifically, each prompt comprises \texttt{\textless{}Model\_Instructions\textgreater{}} for general translation rules and \texttt{\textless{}Task\_Input\textgreater{}} for the task to be translated. The following four stages describe prompt construction; the example stage applies only to one-shot prompting.

\subsubsection{Settings}
\label{sec:workflow-inputs}
At the beginning of each translation task, the program reads the following inputs: (1) \texttt{\{task\}}, containing the task identifier, source entry point, natural-language description, canonical reference implementation, and test assertions; (2) \texttt{\{source\}}, offering the source code; and (3) \texttt{\{target\_name\}}, identifying the target SDK or language. These inputs are used to construct the
\texttt{\textless{}Task\_Input\textgreater{}}
component shown in Figure~\ref{fig:one-shot-program-prompt}.

Then, the configuration identifies the source SDK or language using \texttt{\{source\_name\}} and assigns the sample identifier \texttt{\{SDKsample\_index\}}. The instruction-prefix provider is specified by \texttt{\{prefix\_provider\}}, while SDK version information is supplied through \mbox{\texttt{\{sdk\_versions = \{\}\}}}.
Because the target environments use different file formats and execution interfaces, the configuration branches on the target: for Q\#, it supplies the Q\# file extension \texttt{\{extension\}} and target-specific \texttt{\{runtime\_guidance\}}; for PennyLane, it supplies the corresponding Python file extension and runtime guidance. These branches account for their distinct loading and execution interfaces. The configuration also records the validated source and target implementations in \texttt{\{source\_program\}} and \texttt{\{target\_program\}}, respectively. These SDK identifiers, version constraints, file settings, and runtime instructions are then written as text into the \texttt{\textless{}Model\_Instructions\textgreater{}} component.

\subsubsection{One-shot Example}
The one-shot condition includes one verified source--target translation example; the zero-shot condition omits this component. The example shows how common quantum operations are represented in the target SDK or language. It is intentionally simple and does not reproduce the full functionality of an evaluated task; its purpose is to illustrate the translation conventions while limiting overlap with the task being generated.

The example is supplied through \texttt{\{example\_block\}}. It contains a short gate-implementation example, a source program in \texttt{\{source\_program\}}, and its semantically equivalent target implementation in \texttt{\{target\_program\}}. The paired programs could demonstrate how operations, parameters, bit ordering, control and target roles, and register ownership are expressed across SDKs.


\subsubsection{Model Instructions}
This stage assembles the general translation instructions as text. The \texttt{\{prefix\}} introduces the task, and the text part describes the translation constraints, including the SDKs, versions, runtime guidance, and output format. The model must return a valid JSON object containing \texttt{"target\_code"} for the translated program.
We also include six rules for two purposes: Rules~\texttt{1}--\texttt{5} constrain faithful native execution and assertion-preserving translation, while Rule~\texttt{6} adds an independent statevector check for eligible circuits.

\subsubsection{Task Input}
Task input combines the task-specific information initialized during the Settings stage into the \texttt{\textless{}Task\_Input\textgreater{}} component. It identifies the task and source entry point, specifies the target SDK, and provides the task description, source Qiskit program, and original assertions. Together, these inputs define the concrete translation and its behavioral constraints.

\section{Empirical Study}
\label{sec:study}

\subsection{Research Questions}
We propose three research questions (RQs) to investigate zero-shot translation quality across LLMs and target SDKs, the impact of one-shot prompting, and the failure patterns of the translated programs.

\textbf{RQ1: \RQQuestion{1}}

\textbf{RQ2: \RQQuestion{2}}

\textbf{RQ3: \ReviewChange{\RQQuestion{3}}}

\subsection{Experimental Design}
\label{sec:setup}

\textbf{Benchmark.}  
In the absence of a benchmark specifically designed for quantum code translation, we adopt an open-source benchmark from prior work on evaluating LLM-based quantum code generation, whose overall evaluation framework closely aligns with ours. We select QuanBench~\cite{quanbench}, a recent benchmark of moderate size, for our experiments. Since one canonical reference implementation in QuanBench cannot be reliably reproduced to pass the given assertions, we exclude that task and only use \StudyValue{tasks} of its \StudyValue{tasks.original} original tasks, together with companion tests.

\textbf{LLMs.} 
We group the evaluated LLM configurations into closed-source and open-weight models.
The former group includes \unmarked{\emph{\StudyValue{model.gpt.name} (GPT)}}~\cite{gpt-5-6-sol} configured with medium reasoning effort, while the latter incorporates \emph{\StudyValue{model.pro.name} (Pro)}~\cite{deepseek-pro}, \unmarked{\emph{\StudyValue{model.flash.name} (Flash)}}~\cite{deepseek-flash}, and \emph{\StudyValue{model.glm.name} (GLM)}~\cite{glm-5-2}. All three open-weight models use a temperature of \StudyValue{model.glm.temperature} and sample-specific random seeds \StudyValue{model.glm.seed.first}--\StudyValue{model.glm.seed.last}. GLM uses \emph{\StudyValue{model.glm.reasoning_effort}} reasoning effort. Pro and Flash use \emph{\StudyValue{model.pro.reasoning_plan_effort}} reasoning effort during planning and disable additional reasoning during final code generation. All other model parameters retain their default configurations.

\textbf{Prompt construction.}
\ReviewChange{We translate Qiskit \StudyValue{sdk.qiskit} programs to PennyLane \StudyValue{sdk.pennylane} and to Q\# using Microsoft's Quantum Development Kit (QDK) \StudyValue{sdk.qdk}.}
The zero-shot prompt omits \textbf{One-shot example} in Figure~\ref{fig:one-shot-program-prompt}.
For one-shot prompting, we select Pro and GLM, \ReviewChange{the two configurations with the lowest overall zero-shot performance across the target SDKs} (Section~\ref{sec:rq1}). Their one-shot prompts include one verified source--target translation example for the target SDK. In detail, we select a basic task from Qolumbina~\cite{qolumbina}, a recent benchmark infrastructure of scalable quantum programs, and manually implement semantically equivalent three-qubit programs for this task in Qiskit, PennyLane, and Q\#; \ReviewChange{this choice is intended to reduce the risk of data leakage}.
We choose simple examples that do not duplicate the full functionality of any evaluated task, and these examples cover reversible comparisons, conditional bit flips, and phase marking. 

\textbf{Evaluation metrics.}
(1) \emph{Effectiveness}: Following QuanBench~\cite{quanbench}, we use \pass{$k$} to measure the task-macro correctness of the translated programs, where a single task succeeds if it passes all assertions provided by QuanBench. For each task, we have $n_t$ sample slots (planned translation attempts), among which $c_t$ are successful. Let $\mathcal{T}$ be the set of retained tasks. The metric is defined as
$\pass{$k$}=\frac{1}{|\mathcal{T}|}\sum_{t\in\mathcal{T}}
\left[1-{\binom{n_t-c_t}{k}}/{\binom{n_t}{k}}\right].$
\ReviewChange{In our experiments, we evaluate $k \in \{1, 3\}$ using $n_t = 10$ sample slots for each model--task--target combination.}
(2) \emph{Cost}: We also record the average runtime and output token count per API response to measure the cost of translation.  


\subsection{Empirical Results}
\subsubsection{RQ1}\label{sec:rq1}
\begin{table}[t]
\caption{Zero-shot task-macro \pass{1}/\pass{3} (\%) over \StudyValue{tasks} tasks, with mean generation runtime and output-token use.}
\label{tab:baseline}
\centering\small
\normalcolor
\setlength{\tabcolsep}{2.5pt}
\renewcommand{\arraystretch}{1.18}
\begin{tabular}{lrrrrrr}
\toprule[1pt]
\multicolumn{1}{c}{\textbf{Model}} & \multicolumn{2}{c}{\textbf{Qiskit $\rightarrow$ PennyLane}} & \multicolumn{2}{c}{\textbf{Qiskit $\rightarrow$ Q\#}} & \multicolumn{2}{c}{\textbf{Mean per API response}} \\
\cmidrule(lr){2-3} \cmidrule(lr){4-5} \cmidrule(lr){6-7}
& \multicolumn{1}{c}{\textbf{Pass@1}} & \multicolumn{1}{c}{\textbf{Pass@3}} & \multicolumn{1}{c}{\textbf{Pass@1}} & \multicolumn{1}{c}{\textbf{Pass@3}} & \multicolumn{1}{c}{\textbf{Runtime (s)}} & \multicolumn{1}{c}{\textbf{Output tokens}} \\
\cmidrule(lr){1-1} \cmidrule(lr){2-2} \cmidrule(lr){3-3} \cmidrule(lr){4-4} \cmidrule(lr){5-5} \cmidrule(lr){6-6} \cmidrule(lr){7-7}
GPT & 51.63 & 61.36 & 71.63 & 82.11 & 149.84 & 2,758 \\
GLM & 49.30 & 65.43 & 22.56 & 48.97 & 148.30 & 9,527 \\
Pro & 55.81 & 73.26 & 16.28 & 37.71 & 108.17 & 7,426 \\
Flash & 72.33 & 86.53 & 44.88 & 65.19 & 92.00 & 8,541 \\
\bottomrule[1pt]
\end{tabular}

\end{table}

Across the evaluated zero-shot model--target pairs \ReviewChange{in Table~\ref{tab:baseline}}, \pass{1} ranges from 16.28\% to 72.33\%, indicating substantial variation in translation success: Flash achieves the highest score on PennyLane, while Pro has the lowest on Q\#. The ranking also depends on the target SDK: Flash leads on PennyLane, whereas GPT leads on Q\#. Target effects vary by model as well: GPT performs better on Q\# than on PennyLane, while GLM, Pro, and Flash perform better on PennyLane. In addition, \pass{3} exceeds \pass{1} for every model--target pair, showing that multi-sample evaluation can find passing candidates missed by a single sample. These results suggest that translation performance should be compared for each model and target SDK rather than summarized by a single overall model ranking.
\revised{Means pool responses with complete telemetry across both targets, including unsuccessful attempts; output tokens cover planning and final generation. Flash has the shortest mean runtime (\StudyValue{efficiency.flash.runtime}~s), while GPT uses the fewest output tokens (\StudyValue{efficiency.gpt.output-tokens}). These models also lead in correctness on PennyLane and Q\#, respectively, suggesting that longer generation or greater token use does not necessarily imply higher correctness.}

\subsubsection{RQ2}

\begin{figure}[t]
\centering
\normalcolor
\begin{tikzpicture}
\begin{axis}[width=\columnwidth,height=1.78in,ybar,bar width=4pt,
 ymin=0,ymax=100,ylabel={\pass{k} (\%)},ytick={0,25,50,75,100},
 xtick=data,xticklabels from table={data/prompt-comparison.dat}{group},
 enlarge x limits=0.18,tick label style={font=\scriptsize},
 label style={font=\footnotesize},ymajorgrids=true,grid style={gray!20},
 legend style={font=\scriptsize,legend columns=2,draw=none,at={(0.5,1.02)},anchor=south},
 axis lines*=left]
\addplot[fill=gray!40,draw=black!65] table[x expr=\coordindex,y=baseone]{data/prompt-comparison.dat};
\addplot[fill=blue!55,draw=black!65] table[x expr=\coordindex,y=exampleone]{data/prompt-comparison.dat};
\addplot[fill=gray!80,draw=black!65] table[x expr=\coordindex,y=basethree]{data/prompt-comparison.dat};
\addplot[fill=orange!80,draw=black!65] table[x expr=\coordindex,y=examplethree]{data/prompt-comparison.dat};
\legend{Zero-shot \pass{1},One-shot \pass{1},Zero-shot \pass{3},One-shot \pass{3}}
\end{axis}
\end{tikzpicture}
\caption{Zero-shot versus one-shot prompts over the same \StudyValue{tasks} tasks. PL: PennyLane; QS: Q\#.}
\label{fig:prompt}
\end{figure}
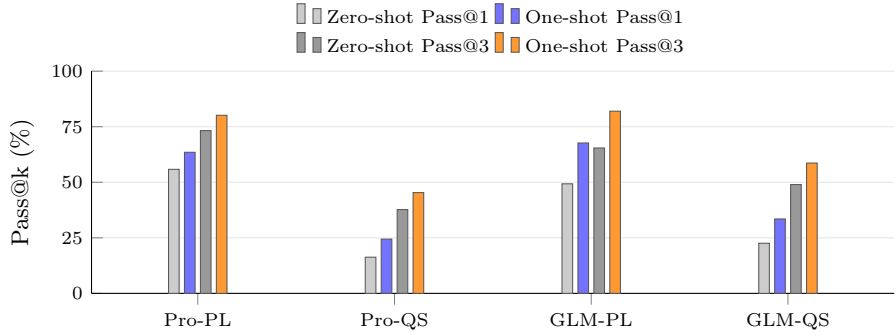

We compare Pro and GLM under the zero-shot and one-shot prompt conditions defined in Section~\ref{sec:setup}, using the same task scope and native evaluation bundle.
\revised{Figure~\ref{fig:prompt} shows that one-shot prompts improve \pass{1} and \pass{3} for both models and target SDKs.} GLM gains \StudyValue{delta.glm.qs.1}--\StudyValue{delta.glm.pl.1} percentage points in \pass{1}, compared with \StudyValue{delta.pro.pl.1}--\StudyValue{delta.pro.qs.1} for Pro. The gains are more pronounced for GLM, particularly on PennyLane, while Pro improves to a similar extent on PennyLane and Q\#. On PennyLane, GLM overtakes Pro under both metrics, showing that model rankings can change with prompt conditions. \revised{Paired examples may help convey SDK-specific mappings and conventions, although this mechanism remains unverified.}


\subsubsection{RQ3}
\begin{table}[t]
\caption{Failure categories under zero-shot and one-shot prompting}
\label{tab:errors}
\centering
\normalcolor
\setlength{\tabcolsep}{2.2pt}
\renewcommand{\arraystretch}{1.10}
\begin{minipage}{\linewidth}
\fontsize{7}{8.4}\selectfont
\begin{tabularx}{\linewidth}{@{}>{\raggedright\arraybackslash}p{0.20\linewidth}>{\raggedright\arraybackslash}Xrrr@{}}
\toprule[1pt]
\multicolumn{1}{c}{\textbf{Failure category}} & \multicolumn{1}{c}{\textbf{Criterion}} & \multicolumn{1}{c}{\textbf{Zero-shot}} & \multicolumn{1}{c}{\textbf{\begin{tabular}[t]{@{}c@{}}Pro + GLM\\One-shot\end{tabular}}} & \multicolumn{1}{c}{\textbf{\begin{tabular}[t]{@{}c@{}}Pro + GLM\\Zero-shot\end{tabular}}} \\
\cmidrule(lr){1-1} \cmidrule(lr){2-2} \cmidrule(lr){3-3} \cmidrule(lr){4-4} \cmidrule(lr){5-5}
Missing source & No source artifact & 700 (39.2\%) & 333 (36.7\%) & 492 (44.7\%) \\
SDK compilation & Target code not compilable & 624 (34.9\%) & 349 (38.5\%) & 407 (37.0\%) \\
Assertion rejection & Test assertion violated & 160 (9.0\%) & 81 (8.9\%) & 53 (4.8\%) \\
Candidate execution & Load, API, or runtime failure & 143 (8.0\%) & 98 (10.8\%) & 79 (7.2\%) \\
SDK incompatibility & Interface/\allowbreak{}dependency-related traceback & 160 (9.0\%) & 46 (5.1\%) & 70 (6.4\%) \\
Total & All failed samples & 1,787 (100.0\%) & 907 (100.0\%) & 1,101 (100.0\%) \\
\bottomrule[1pt]
\end{tabularx}

\end{minipage}
\end{table}
We evaluate \StudyValue{models.count} LLM configurations on \StudyValue{targets.count} target SDKs and \StudyValue{tasks} tasks, with \StudyValue{samples.per-task} sample slots per combination, yielding $\StudyValue{models.count} \times \StudyValue{targets.count} \times \StudyValue{tasks} \times \StudyValue{samples.per-task} = \text{\StudyValue{baseline.slots}}$ zero-shot sample slots. Among these, \StudyValue{baseline.passed} pass and \StudyValue{baseline.failed} fail (Table~\ref{tab:errors}).
\revised{Table~\ref{tab:errors} reports mutually exclusive failure categories, their criteria, counts, and within-column failure percentages for all zero-shot models and the Pro + GLM prompt comparison. Scripts automatically classify failed samples using generation and evaluation statuses and tracebacks. Missing source (\StudyValue{failures.missing_generation_artifact}) and SDK compilation (\StudyValue{failures.sdk_compilation_error}) dominate, identifying artifact delivery and compilation as the main observed barriers. Other categories indicate execution or interface problems without establishing algorithmic incorrectness.}

\revised{For Pro and GLM across both targets, failed samples decrease from \StudyValue{failures.without_program_prompt.total} under zero-shot prompting to \StudyValue{failures.with_program_prompt.total} with one-shot prompting. Missing source, SDK compilation, and SDK incompatibility decrease, while assertion rejection and candidate execution increase. SDK compilation becomes the largest one-shot category (\StudyValue{failures.with_program_prompt.sdk_compilation_error.share}\%), while missing source dominates under zero-shot prompting (\StudyValue{failures.without_program_prompt.missing_generation_artifact.share}\%).}


\section{Conclusion and Future Work}
\label{sec:conclusion}
Our study presents an initial proposal and a pilot evaluation of LLM-based quantum code translation, with four LLM configurations, \ReviewChange{two translation targets (PennyLane and Q\#)}, 43 benchmark tasks, and a comparison between zero-shot and one-shot prompts. The results show substantial variation in translation quality across the evaluated LLM configurations and target SDKs, with zero-shot \pass{1} ranging from 16.28\% to 72.33\%.
The most common failure category is missing source, followed by SDK compilation.


\label{sec:future}
In future work, we plan to build a dedicated cross-SDK translation benchmark by integrating existing quantum programs, including scalable programs such as Qolumbina~\cite{qolumbina}, with paired source--target implementations and executable tests across SDKs and program sizes. We will also develop translation-specific metrics and assertions for semantic preservation and required structural consistency, drawing on quantum software testing~\cite{li2026methodological}. These checks will cover relative phase, bit ordering, and probabilistic outputs to complement output-based evaluation. Finally, we will conduct broader empirical studies of prompting strategies and explore execution-guided repair while assessing translation quality and cost.

\section*{Data Availability}
The source code, experimental artifacts, and reproduction instructions are available in an anonymized repository.\footnote{\url{https://anonymous.4open.science/r/quantum-translation-artifact-2A7C/}}
\clearpage
\bibliographystyle{splncs04}
\bibliography{components/ref}
\end{document}